\documentclass[10pt]{article}
\usepackage{graphicx}
\usepackage{longtable}
\usepackage{times}
\usepackage{setspace}
\usepackage{dcolumn}
\usepackage{bm}
\usepackage{rotating}
\usepackage{hyperref}
\usepackage{epstopdf}
\begin{document}
\begin{center}
    \textbf{The change in the direction of the electric field using the Lorentz-type transformation}
\end{center}
\begin{center}
    $ Anuj \ Kumar \ Dubey^{1} \  and \ Tanay \ Ghosh^{2}$
\end{center}
\begin{center}
    \textit{1. Department of Applied Science, Allenhouse Institute of Technology,  \\ Kanpur, Uttar Pradesh, India \\  E-mail:danuj67@gmail.com} \\
\end{center}
\begin{center}
\begin{center}
    \textit{2. Department of Science, Sunbeam Women's College, Varuna  \\ Varansi, Uttar Pradesh, India \\  E-mail:$tanay\_ghosh$@hotmail.co.uk} \\
\end{center}
\end{center}
\begin{abstract}
 In the present work, we will derive the expression for the change in direction of the electric field using the Lorentz transformation for inertial frame or un-accelerated frame. The acceleration also influences the observed rate of a moving clock in addition to the influence due to its velocity. So we will also derive the expression for the change in direction of the electric field using the Lorentz-type transformation (as given by Friedman) for non-inertial frame or uniformly accelerated frame.
\end{abstract}
\section{Introduction}\label{section1}
According to Einstein's special theory of relativity time is a relative notion, which means that the observed rate of a moving clock differs from the rate of a stationary clock. In fact, acceleration also influences the observed rate of a moving clock in addition to the influence due to its velocity. Today, there are several existing experimental techniques to test whether acceleration influences the observed rate of a clock.

Mashhoon and Muench \cite{1} had investigated the limitations of length measurements by accelerated observers in Minkowski spacetime that follow from the hypothesis of locality (the assumption that an accelerated observer at each instant is equivalent to an otherwise identical momentarily comoving inertial observer).

Caianiello \cite{2} had predicted the existence of a maximal acceleration for massive objects based on the time-energy uncertainty relation in quantum mechanics (see Papini and Wood \cite{3}, Papini et al. \cite{4}).

Friedman and Gofman \cite{5,6,7} had shown that there are only two possible types of transformations between uniformly accelerated systems. The first allowable type of transformation holds if and only if the Clock Hypothesis is true. If the Clock Hypothesis is not true, the transformation is of Lorentz-type and implies the existence of a universal maximal acceleration. They used the proper velocity-time description of events rather than the usual space-time description in order to obtain the linear transformations. They derived the  Lorentz-type transformations between uniformly accelerated systems in extended relativity. Friedman \cite{7} showed that extended relativity predicts that there is a maximal acceleration of the order of $10^{19} \frac{m}{s^{2}}$. The Friedman's extended relativity predicts that the speed of any moving object is limited by speed of light (c) and the magnitude of its acceleration is limited by a maximal acceleration ($a_{m}$).

Friedman and Scarr \cite{8} had used the generalized Fermi-Walker transport to construct a one-parameter family of inertial frames which are instantaneously comoving to a uniformly accelerated observer. They showed that the solutions of uniformly accelerated motion have constant acceleration in the comoving frame. Assuming the weak hypothesis of locality, they obtained local spacetime transformations from a uniformly accelerated frame to an inertial frame. They also showed that the spacetime transformations between two uniformly accelerated frames with the same acceleration are Lorentz. They also computed the metric at an arbitrary point of a uniformly accelerated frame.

Friedman and Scarr \cite{9}, had obtained an equation for uniform acceleration in a general curved spacetime from linear acceleration to the full Lorentz covariant uniform acceleration. They used generalized Fermi-Walker transport to parallel transport the Frenet basis along the trajectory. In flat spacetime, they obtained velocity and acceleration transformations from a uniformly accelerated system to an inertial system. They also obtained the time dilation between accelerated clocks. They applied the obtained acceleration transformations to the motion of a charged particle in a constant electromagnetic field and recovered the Lorentz-Abraham-Dirac equation. 

With this background the present  paper is organized as follows:

In section-\ref{section2}, we will derive the expression for the change in direction of the electric field using the Lorentz transformation in inertial frame (un-accelerated frame).
In section-\ref{section3}, we  derive the expression for Lorentz type transformation in non-inertial frame (uniformly accelerated frame.  In section-\ref{section4}, we  derive the expression for the change in direction of the electric field using the Lorentz type transformation in non-inertial frame (uniformly accelerated frame). Finally, some conclusions are made in section-\ref{section5}.

\section{The change in the direction of the electric field using the Lorentz transformation in inertial frame}\label{section2}
It should be noted that to avoid confusion, we will always use V to signify the constant relative velocity of two inertial systems, and v for the velocity of a moving particle.\\

The Lorentz transformation formula from one inertial frame of reference (K) to another inertial frame of reference $(K^{'})$, which is moving with constant relative velocity V along the x-axis, can we expressed as (page 496 of Griffiths \cite{10} and page 11 of Landau and Lifshitz \cite{11}):
 \begin{equation}\label{28}
   t= \frac{t^{'}+ \frac{V}{c^{2}}x^{'}}{{\sqrt{1-\frac{V^{2}}{c^{2}}}}}
\end{equation}
\begin{equation}\label{29}
   x=\frac{x^{'}+ V t^{'}}{{\sqrt{1-\frac{V^{2}}{c^{2}}}}}
\end{equation}
\begin{equation}\label{30}
   y=y^{'}
\end{equation}
\begin{equation}\label{31}
   z=z^{'}
\end{equation}
The Lorentz reverse transformation formula can be written as:
 \begin{equation}\label{32}
   t^{'}=\frac{t- \frac{V}{c^{2}}x}{{\sqrt{1-\frac{V^{2}}{c^{2}}}}}
\end{equation}
\begin{equation}\label{33}
   x^{'}=\frac{x-V t}{{\sqrt{1-\frac{V^{2}}{c^{2}}}}}
\end{equation}
\begin{equation}\label{34}
   y^{'}=y
\end{equation}
\begin{equation}\label{35}
    z^{'}=z
\end{equation}

The Lorentz transformation formula for four vectors $(x^{\mu}= ct, x, y, z)$ from one inertial frame of reference (K) to another inertial frame of reference $(K^{'})$, which is moving with constant relative velocity V along the x-axis, can we expressed in matrix form as (page 500 of Griffiths \cite{10}):
\begin{equation}\label{36}
\left(
            \begin{array}{c}
              x^{'0}(=ct^{'}) \\
              x^{'1}=(x^{'}) \\
              x^{'2}=(y^{'}) \\
              x^{'3}=(z^{'}) \\
   \end{array}
          \right)
          = \left(
  \begin{array}{cccc}
    \gamma &  -\gamma \beta &  0 &  0  \\
     -\gamma \beta   &  \gamma &  0 &  0  \\
     0 &  0 &  1 &  0 \\
       0 &  0 &  0 &  1 \\
  \end{array}
\right)\left(
            \begin{array}{c}
             x^{0}(=ct) \\
              x^{1}(=x) \\
              x^{2}(=y) \\
              x^{3}(=z) \\
            \end{array}
          \right)
\end{equation}

Here
\begin{equation}\label{37}
\gamma=\frac{1}{\sqrt{1-\frac{V^{2}}{c^{2}}}},
\end{equation}
\begin{equation}\label{38}
\beta =\frac{V}{c},
\end{equation}
and
\begin{equation}\label{39}
\Lambda^{\mu}_{\nu} = \left(
  \begin{array}{cccc}
    \gamma &  -\gamma \beta &  0 &  0  \\
     -\gamma \beta   &  \gamma &  0 &  0  \\
     0 &  0 &  1 &  0 \\
       0 &  0 &  0 &  1 \\
  \end{array}
\right)
\end{equation}
is the Lorentz transformation matrix in equation (\ref{36})(the superscript $\mu$ labels the row, the subscript $\nu$ labels the column).

Letting Greek indices run from 0 to 3, the above equation (\ref{36}) can be distilled into a single equation as:
\begin{equation}\label{40}
   x^{'\mu}=\Sigma \  (\Lambda^{\mu}_{\nu}) \ x^{\nu}
 \end{equation}
The electromagnetic field tensor $F^{\mu\nu}$ is given as (page 531 of Griffiths \cite{10}):
\begin{equation}\label{41}
F^{\mu \nu} = \left(
  \begin{array}{cccc}
    0 &  \frac{E_{x}}{c} &  \frac{E_{y}}{c} &  \frac{E_{z}}{c}  \\
     -\frac{E_{x}}{c}   &  0 &  B_{z} &  -B_{y}  \\
    - \frac{E_{y}}{c} & -B_{z} &  0 &  B_{x} \\
      - \frac{E_{z}}{c} &  B_{y} &  -B_{x} &  0 \\
  \end{array}
\right)
\end{equation}
We can write the transformation as:
\begin{equation}\label{42}
 F^{'\mu \nu}= \Lambda^{\mu}_{\rho} \Lambda^{\nu}_{\sigma} F^{\rho \sigma}
\end{equation}
We can write the above transformation equation (\ref{42}) in matrix form as  $F^{'}=\Lambda F \Lambda^{T}$:
\begin{equation}\label{43}
F^{'\mu \nu} =\left(
 \begin{array}{cccc}
    \gamma &  -\gamma \beta &  0 &  0  \\
     -\gamma \beta   &  \gamma &  0 &  0  \\
     0 &  0 &  1 &  0 \\
       0 &  0 &  0 &  1 \\
  \end{array}
\right) \left(
 \begin{array}{cccc}
    0 &  \frac{E_{x}}{c} &  \frac{E_{y}}{c} &  \frac{E_{z}}{c}  \\
    -\frac{E_{x}}{c}   &  0 &  B_{z} &  -B_{y}  \\
    - \frac{E_{y}}{c} & -B_{z} &  0 &  B_{x} \\
      - \frac{E_{z}}{c} &  B_{y} &  -B_{x} &  0 \\
  \end{array}
\right)
\left(
  \begin{array}{cccc}
   \gamma &  -\gamma \beta &  0 &  0  \\
     -\gamma \beta   &  \gamma &  0 &  0  \\
    0 &  0 &  1 &  0 \\
       0 &  0 &  0 &  1 \\
  \end{array}
\right)
\end{equation}
After simplification above equation (\ref{43}) becomes:
$$F^{'\mu \nu}=\left(
 \begin{array}{cccc}
    0 &  \frac{E^{'}_{x}}{c} &  \frac{E^{'}_{y}}{c} &  \frac{E^{'}_{z}}{c}  \\
    -\frac{E^{'}_{x}}{c}   &  0 &  B^{'}_{z} &  -B^{'}_{y}  \\
    - \frac{E^{'}_{y}}{c} & -B^{'}_{z} &  0 &  B^{'}_{x} \\
      - \frac{E^{'}_{z}}{c} &  B^{'}_{y} &  -B^{'}_{x} &  0 \\
  \end{array}
\right)=$$
\begin{equation}\label{44}
\left(
 \begin{array}{cccc}
    0 &  \frac{E_{x}}{c} & \frac{\gamma}{c} (E_{y}-V B_{z})   &  \frac{\gamma}{c} (E_{y}+V B_{y})\\
    -\frac{E_{x}}{c}   &  0 &  \gamma (B_{z}-\frac{V}{c^{2}}E_{y}) & - \gamma (B_{y}+\frac{V}{c^{2}}E_{z})  \\
    - \frac{\gamma}{c} (E_{y}-V B_{z}) & -\gamma (B_{z}-\frac{V}{c^{2}}E_{y}) &  0 &  B_{x} \\
      - \frac{\gamma}{c} (E_{z}+V B_{y}) & \gamma (B_{y}+\frac{V}{c^{2}}E_{z}) &  -B_{x} &  0 \\
  \end{array}
\right)
\end{equation}
From above equation, now we can express the components of antisymmetric second rank electromagnetic field tensor $(F^{\mu \nu})$ in terms of the components of the electric field $(\mathbf{E})$, and magnetic field (B) from one inertial frame of reference (K) to another inertial frame of reference $(K^{'})$, which is moving with constant relative velocity V along the x-axis.

Using the above equation (\ref{44}), the transformation formulas for the components of the electric field $(\mathbf{E})$ can be written as:
\begin{equation}\label{45}
   E_{x}=E^{'}_{x}
 \end{equation}
 \begin{equation}\label{46}
   E_{y}= \gamma (E^{'}_{y}+V B^{'}_{z})
 \end{equation}
 \begin{equation}\label{47}
   E_{z}= \gamma (E^{'}_{z}-V B^{'}_{y})
 \end{equation}
Similarly  the transformation formulas for the components of the magnetic field $(\mathbf{E})$ can be written as:
\begin{equation}\label{48}
   B_{x}=B^{'}_{x}
 \end{equation}
 \begin{equation}\label{49}
   B_{y}=\gamma (B^{'}_{y} - \frac{V}{c^{2}} E^{'}_{z})
 \end{equation}
 \begin{equation}\label{50}
   B_{z}=\gamma (B^{'}_{z} +  \frac{V}{c^{2}}E^{'}_{y})
 \end{equation}
Thus the electric and magnetic fields, like the majority of physical quantities, are relative; that their properties are different in different reference systems. In particular, the electric and magnetic field can be equal to zero in one reference system and at the same time be present in another system (page 67 of Landau and Lifshitz \cite{11}).

Substituting $B^{'}_{z}=\frac{E^{'}_{z}}{c}$ and $B^{'}_{y}=\frac{E^{'}_{y}}{c}$, equations (\ref{49}) and (\ref{50}) becomes;
\begin{equation} \label{51}
   E_{y}= \gamma (E^{'}_{y}+\frac{V}{c} E^{'}_{z})
 \end{equation}
 \begin{equation}\label{52}
   E_{z}= \gamma (E^{'}_{z}-\frac{V}{c} E^{'}_{y})
 \end{equation}
 When light is emitted from the source, if we assume in the source frame that, the electric vector is making an angle $45^o$ with the y axis then $E^{'}_{y}=E^{'}_{z}$.
After substituting $E^{'}_{y}=E^{'}_{z}$, equations (\ref{51}) and (\ref{52}) becomes;
\begin{equation}\label{53}
   E_{y}= \gamma E^{'}_{y}(1+\frac{V}{c})
 \end{equation}
 \begin{equation}\label{54}
   E_{z}= \gamma E^{'}_{y}(1-\frac{V}{c})
 \end{equation}
From above equations (\ref{53}) and (\ref{54}), we can write;
\begin{equation}\label{55}
\theta_{E}= tan^{-1} (\frac{E_{z}}{E_{y}})=tan^{-1} (\frac{1-\frac{V}{c}}{1+\frac{V}{c}})
 \end{equation}
If relative velocity between the two inertial frames (V) equals to zero or V $\ll$ c, then $\theta_{E}=45^o$. So now we can introduce the angle $\Delta \theta_{E}$, which may be considered as the change in direction of the electric field on transforming one inertial frame of reference to another inertial frame of reference and given as:
\begin{equation}\label{56}
\Delta \theta_{E}=tan^{-1}  (\frac{1-\frac{V}{c}}{1+\frac{V}{c}})-45^o
\end{equation}
\section{Lorentz type transformation in non-inertial frame}\label{section3}
As we know that the relation between the dilated time (t) and proper time $\tau$ is $t=\frac{\tau}{\sqrt{1-\frac{V^{2}}{c^{2}}}}$. Now we can introduce a new proper velocity-time description of events for describing the transformations between two uniformly accelerated systems. Proper velocity (p-velocity in short) of an object is the derivative of the object's velocity with respect to the proper time $(\tau)$. Proper velocity $(\mathbf{u})$ of an object moving with uniform velocity $(\mathbf{v})$ is defined as:
\begin{equation}\label{57}
  \mathbf{u}=\frac{d\mathbf{r}}{d\tau}=\frac{d\mathbf{r}}{dt} \frac{d\mathbf{t}}{d\tau}=\mathbf{v} (\frac{1}{\sqrt{1-\frac{V^{2}}{c^{2}}}})
\end{equation}

The proper velocity is also the canonically conjugate variable to the position in the relativistic phase-space.

The relativistic acceleration \lq $\mathbf{a}$\rq, \ is the derivative of p-velocity with respect to time t and can be written as (\cite{12} p.71):
\begin{equation}\label{58}
  \mathbf{a}=  \frac{d\mathbf{u}}{dt}
\end{equation}
If an object moves with constant proper acceleration, then its p-velocity satisfies the equation,
\begin{equation}\label{59}
 \frac{d^{2}\mathbf{u}}{dt^{2}}=0
\end{equation}
We will say that an object is uniformly accelerated if its proper acceleration is constant,
or equivalently, satisfies the above equation (\ref{59}). If the velocity of a uniformly accelerated object is parallel to the acceleration, then it moves with the well-known hyperbolic motion (see \cite{12,13,14}).

Following works of Friedman (\cite{5,6,7}), the Lorentz-type proper velocity-time transformation formula from one non-inertial frame of reference (K) to another non-inertial frame of reference $(K^{'})$, which is moving with constant relative acceleration (a) along the x-axis can be written as:
 \begin{equation}\label{60}
   t= \tilde{\gamma} (t^{'}+\frac{a u^{'}_{x}}{a^{2}_{m}})
 \end{equation}
 \begin{equation}\label{61}
   u_{x}= \tilde{\gamma} (at^{'}+u^{'}_{x})
 \end{equation}
 \begin{equation}\label{62}
   u_{y}=u^{'}_{y}
 \end{equation}
 \begin{equation}\label{63}
   u_{z}= u^{'}_{z}
 \end{equation}
 Here, $\tilde{\gamma} = \frac{1}{\sqrt{1-\frac{a^{2}}{a^{2}_{m}}}}$ is the time dilation due to acceleration.

 The clock hypothesis in Einstein (1911) states that the rate of an accelerated clock is equal to that of a comoving un-accelerated clock. If the clock hypothesis is valid, then the transformations between uniformly accelerated systems are Galilean. If not, these transformations are of Lorentz-type, and, moreover, there exists a universal maximal acceleration, which we denote by $a_{m}$.

Friedman and Gofman \cite{5} predicted that there is a maximal acceleration of the order of $10^{19} \frac{m}{s^{2}}$. The form $\tilde{ds^{2}} =  a^{2}_{m} dt^{2}-du^{2}$ is an invariant for uniformly accelerated systems. This implies that the acceleration a of any massive object is limited by the maximal one $a<a_{m}$. Moreover, the proper velocity time trajectory of such objects is inside the \lq light cone\rq \ $u<a_{m} t$ in the proper velocity time continuum \cite{7}.

 The Lorentz-type proper velocity-time transformation formulas (\ref{60}) to (\ref{63}) from one non-inertial frame of reference (K) to another non-inertial frame of reference $(K^{'})$, which is moving with constant relative acceleration (a) along the x-axis in the matrix form can be written as:
\begin{equation}\label{64}
\left(
            \begin{array}{c}
             a_{m}t^{'} \\
              u_{x}^{'} \\
             u_{y}^{'} \\
              u_{z}^{'} \\
   \end{array}
          \right)
          = \left(
  \begin{array}{cccc}
     \tilde{\gamma} &  - \frac{\tilde{\gamma} a}{a_{m}} &  0 &  0  \\
    - \frac{\tilde{\gamma} a}{a_{m}}  &  \tilde{\gamma}  &  0 &  0  \\
     0 &  0 &  1 &  0 \\
       0 &  0 &  0 &  1 \\
  \end{array}
\right)\left(
            \begin{array}{c}
            a_{m} t \\
             u_{x}  \\
             u_{x}  \\
             u_{x} \\
            \end{array}
          \right)
\end{equation}
 From above equation (\ref{64}), we can write the Lorentz-type proper velocity-time transformation matrix $(S^{\mu}_{\nu})$ as given below:
  \begin{equation}\label{65}
S^{\mu}_{\nu}= \left(
  \begin{array}{cccc}
    \tilde{\gamma} &   -\frac{\tilde{\gamma} a}{a_{m}} &  0 &  0  \\
    - \frac{\tilde{\gamma} a}{a_{m}}  &  \tilde{\gamma}  &  0 &  0  \\
     0 &  0 &  1 &  0 \\
       0 &  0 &  0 &  1 \\
  \end{array}
\right)
\end{equation}
\section{The change in the direction of the electric field using the Lorentz type transformation in non-inertial frame}\label{section4}
Now we can write the transformation equation in matrix form for the antisymmetric second
rank electromagnetic field tensor ($F^{\mu \nu}$), as  $F^{'}= S F S^{T}$:
\begin{equation}\label{66}
F^{'\mu \nu} =\left(
 \begin{array}{cccc}
  \tilde{\gamma} &   -\frac{\tilde{\gamma} a}{a_{m}} &  0 &  0  \\
    - \frac{\tilde{\gamma} a}{a_{m}}  &  \tilde{\gamma}  &  0 &  0  \\
     0 &  0 &  1 &  0 \\
       0 &  0 &  0 &  1 \\
  \end{array}
\right) \left(
 \begin{array}{cccc}
    0 &  \frac{E_{x}}{c} &  \frac{E_{y}}{c} &  \frac{E_{z}}{c}  \\
    -\frac{E_{x}}{c}   &  0 &  B_{z} &  -B_{y}  \\
    - \frac{E_{y}}{c} & -B_{z} &  0 &  B_{x} \\
      - \frac{E_{z}}{c} &  B_{y} &  -B_{x} &  0 \\
  \end{array}
\right)
\left(
  \begin{array}{cccc}
  \tilde{\gamma} & -  \frac{\tilde{\gamma} a}{a_{m}} &  0 &  0  \\
   -  \frac{\tilde{\gamma} a}{a_{m}}  &  \tilde{\gamma}  &  0 &  0  \\
     0 &  0 &  1 &  0 \\
       0 &  0 &  0 &  1 \\
  \end{array}
\right)
\end{equation}
After simplification above equation (\ref{66}) becomes:
$$F^{'\mu \nu}=\left(
 \begin{array}{cccc}
    0 &  \frac{E^{'}_{x}}{c} &  \frac{E^{'}_{y}}{c} &  \frac{E^{'}_{z}}{c}  \\
    -\frac{E^{'}_{x}}{c}   &  0 &  B^{'}_{z} &  -B^{'}_{y}  \\
    - \frac{E^{'}_{y}}{c} & -B^{'}_{z} &  0 &  B^{'}_{x} \\
      - \frac{E^{'}_{z}}{c} &  B^{'}_{y} &  -B^{'}_{x} &  0 \\
  \end{array}
\right)=$$
\begin{equation}\label{67}
\left(
 \begin{array}{cccc}
    0 &  \frac{E_{x}}{c} & \tilde{\gamma}(\frac{E_{y}}{c}-B_{z}\frac{a}{a_{m}}) &  \tilde{\gamma}(\frac{E_{z}}{c}+B_{y}\frac{a}{a_{m}})  \\
    -\frac{E_{x}}{c}  &  0 &  \tilde{\gamma} (B_{z}- \frac{E_{y}}{c} \frac{a}{a_{m}})&  -\tilde{\gamma}(B_{y}+ \frac{E_{z}}{c} \frac{a}{a_{m}})  \\
    - \tilde{\gamma}(\frac{E_{y}}{c}-B_{z}\frac{a}{a_{m}}) & -\tilde{\gamma} (B_{z}-\frac{E_{y}}{c} \frac{a}{a_{m}}) &  0 &  B_{x} \\
     - \tilde{\gamma}(\frac{E_{z}}{c}+B_{y}\frac{a}{a_{m}}) &  \tilde{\gamma}(B_{y}+ \frac{E_{z}}{c} \frac{a}{a_{m}}) &  -B_{x} &  0 \\
  \end{array}
\right)
\end{equation}
Now using the above equation (\ref{67}), we can express the components of antisymmetric second rank electromagnetic field tensor $(F^{\mu \nu})$ in terms of the components of the electric field $(\mathbf{E})$, and magnetic field (B) from one non-inertial frame of reference (K) to another non-inertial frame of reference $(K^{'})$, which is moving with constant relative acceleration (a) along the x-axis.
The transformation formulas for the components of the electric field $(\mathbf{E})$ can be written as:
\begin{equation}\label{68}
   E_{x}=E^{'}_{x}
 \end{equation}
 \begin{equation}\label{69}
   E_{y}=\tilde{\gamma}(E^{'}_{y}+ B^{'}_{z}c \frac{a}{a_{m}})
 \end{equation}
 \begin{equation}\label{70}
   E_{z}=\tilde{\gamma}(E^{'}_{z}- B^{'}_{y}c \frac{a}{a_{m}})
 \end{equation}

Similarly  the transformation formulas for the components of the magnetic field $(\mathbf{E})$ can be written as:
\begin{equation}\label{71}
   B_{x}=B^{'}_{x}
 \end{equation}
 \begin{equation}\label{72}
   B_{y}=\tilde{\gamma}(B^{'}_{y} - \frac{E^{'}_{z}}{c} \frac{a}{a_{m}})
 \end{equation}
 \begin{equation}\label{73}
   B_{z}=\tilde{\gamma}(B^{'}_{z} +\frac{E^{'}_{y}}{c} \frac{a}{a_{m}})
 \end{equation}
 Substituting $B^{'}_{z}=\frac{E^{'}_{z}}{c}$ and $B^{'}_{y}=\frac{E^{'}_{y}}{c}$, equations (\ref{69}) and (\ref{70}) becomes;
\begin{equation}\label{74}
   E_{y}=\tilde{\gamma}(E^{'}_{y}+ E^{'}_{z} \frac{a}{a_{m}})
 \end{equation}
 \begin{equation}\label{75}
E_{z}=\tilde{\gamma}(E^{'}_{z}- E^{'}_{y}c \frac{a}{a_{m}})
 \end{equation}
 When light is emitted from the source, if we assume in the source frame that, the electric vector is making an angle $45^o$ with the y axis then $E^{'}_{y}=E^{'}_{z}$.
After substituting $E^{'}_{y}=E^{'}_{z}$, above equations (\ref{74}) and (\ref{75}) becomes;
\begin{equation}\label{76}
   E_{y}=\tilde{\gamma}E^{'}_{y}(1+\frac{a}{a_{m}})
 \end{equation}
 \begin{equation}\label{77}
 E_{z}=\tilde{\gamma}E^{'}_{y}(1-\frac{a}{a_{m}})
 \end{equation}
From above equations (\ref{76}) and (\ref{77}), we can write;
\begin{equation}\label{78}
\tilde{\theta}_{E}= tan^{-1} (\frac{E_{z}}{E_{y}})=tan^{-1} (\frac{1-\frac{a}{a_{m}}}{1+\frac{a}{a_{m}}})
 \end{equation}
If relative acceleration (a) between the two non-inertial frames (a) equals to zero or $a \ll a_{m}$ , then $\tilde{\theta}_{E}=45^o$. So now we can introduce the angle $\Delta \tilde{\theta}_{E}$, which may be considered as the change in direction of the electric field on transforming from one non-inertial frame of reference to another non-inertial frame of reference and given as:
\begin{equation}\label{79}
\Delta \tilde{\theta}_{E}=tan^{-1}  (\frac{1-\frac{a}{a_{m}}}{1+\frac{a}{a_{m}}})-45^o
\end{equation}
\section{Conclusions}\label{section5}
In the present work, we found the expression for the change in the direction of the electric field using the Lorentz transformation from one inertial frame of reference to another inertial frame of reference as (earlier obtained expression (\ref{56})):
$$\Delta \theta_{E}=tan^{-1}  (\frac{1-\frac{V}{c}}{1+\frac{V}{c}})-45^o$$
If relative velocity between the two inertial frames (V) equals to zero or V $\ll$ c, then $\theta_{E}=45^o$.

We also found the expression for the change in direction of the electric field using the Lorentz type transformation from one non inertial frame of reference to another non inertial frame of reference as (earlier obtained expression (\ref{79})):
$$\Delta \tilde{\theta}_{E}=tan^{-1}  (\frac{1-\frac{a}{a_{m}}}{1+\frac{a}{a_{m}}})-45^o$$
If relative acceleration (a) between the two non-inertial frames (a) equals to zero or $a \ll a_{m}$ , then $\tilde{\theta}_{E}=45^o$.
\section{Acknowledgement}
A K Dubey wish to thank Dr. Atul Kumar Agnihotri, Prof. Somendra Shukla and Prof. Bhagwan Jagwani, for inspiring discussions and also for their support and providing the research facilities at Allenhouse Institute of Technology, Kanpur.


\begin{thebibliography}{99}
\bibitem{1} B.  Mashhoon and U. Muench: Length measurement in accelerated systems, Ann. Phys., Lpz., 11, 53247 (2002).
\bibitem{2} E. R. Caianiello: Lett. Nuovo Cimento, 32, 65 (1981).
\bibitem{3} G. Papini and  W. R. Wood: Phys. Lett. A, 170, 409 (1992).
\bibitem{4} G. Papini, G. Scarpetta, V. Bozza, A. Feoli and G. Lambiase: Phys. Lett. A, 300, 603 (2002).
\bibitem{5} Y. Friedman and Y. Gofman: A new relativstic kinematics of accelerated systems, Phys. Scr. 82, 015004 (2010). arXiv:gr-qc/0509004v2
\bibitem{6} Y. Friedman: Extending the relativity of time, Journal of Physics: Conference Series, 437, 012017 (2013).
\bibitem{7} Y. Friedman: The maximal acceleration, Extended Relativistic Dynamics and Doppler type shift for an accelerated source, arXiv:0910.5629v3 [physics.class-ph](2010).
\bibitem{8} Y. Friedman and T. Scarr: Spacetime transformations from a uniformly accelerated frame, Phys. Scr. 87, 055004 (2013).
\bibitem{9} Y. Friedman and T. Scarr: Uniform accelerations in general relativity, Gen. Relativ. Gravit. 47, 121 (2015).
\bibitem{10} D. J. Griffiths: Introduction to Electrodynamics, 3rd edn., Prentice-Hall, Inc. p. 496, 500, (1999).
\bibitem{11} L. D. Landau and E. M. Lifshitz: The Classical Theory of Fields, Vol. 2, 4th revised edn., Butterworth- Heinemann, Indian Reprint, p. 11, 15, 66 (2008).
 \bibitem{12} W. Rindler:  Relativity: Special, General, and Cosmological, Oxford University Press, New-York (2001).
\bibitem{13} J. Franklin: Eur. J. Phys. 31, 291 (2010).
\bibitem{14} C. Moller: The theory of Relativity, Clarendon Press, Oxford (1972).
\end{thebibliography}
\end{document}